\documentclass[onecolumn,aps,superscriptaddress,longbibliography,notitlepage]{revtex4-1}

\usepackage{graphicx}
\usepackage[normalem]{ulem}
\usepackage{color}

\usepackage{latexsym}
\usepackage[caption=false]{subfig}

\usepackage[utf8]{inputenc}

\usepackage{amsmath,amsfonts,amssymb}
\usepackage{braket}
\usepackage{hyperref}
\hypersetup{colorlinks,linkcolor={blue},citecolor={blue},urlcolor={red}}

\begin{document}

\title{Dynamics of single-mode nonclassicalities and quantum correlations in the Jaynes-Cummings model}
\author{Sriram Akella}\thanks{asriram@cmi.ac.in}
\affiliation{Chennai Mathematical Institute, H1, Sipcot IT Park, Siruseri, Kelambakkam, Tamilnadu, 603103, India}
\affiliation{Department of Theoretical Physics, Tata Institute of Fundamental
Research, Homi Bhabha Road, Colaba, Mumbai 400005, India}

\author{Kishore Thapliyal} \thanks{kishore.thapliyal@upol.cz}
\affiliation{Joint Laboratory of Optics of
Palack\'{y} University and Institute of Physics of CAS, Faculty of Science,
Palack\'{y} University, 17. listopadu 12, 771 46 Olomouc, Czech Republic}

\author{H S Mani} \thanks{hsmani@gmail.com }
\affiliation{Chennai Mathematical Institute, H1, Sipcot IT Park, Siruseri, Kelambakkam, Tamilnadu, 603103, India}

\author{Anirban Pathak} \thanks{anirban.pathak@gmail.com, Corresponding author}
\affiliation{Jaypee Institute of Information Technology, A-10, Sector-62, Noida, UP-201309, India}

\begin{abstract}
 Dynamics of atom-field correlations and single-mode nonclassicalities present in the resonant Jaynes-Cummings model are investigated using negativity and entanglement potential for a set of initial states. The study has revealed the interplay between three different types of nonclassicality present in the model and established that the nonclassicality is continuously exchanged between the field and atom through the atom-field correlations. Further, it is observed that the entanglement potential does not capture all the single-mode nonclassicality and there exists some residual nonclassicality in the reduced single-mode states at the output of the beam splitter which is not captured by the entanglement in which single-mode nonclassicality is quantitatively mapped in Asboth's criterion. Additional layers of beam splitters are added to deplete all the nonclassicality and to reveal that almost all the residual nonclassicality is captured with three layers of beam splitters. Further, the reduced states of the atom and field have
zero (non-zero) quantum coherence in the Fock basis when the atom-field correlations are maximum if the field (or atom) has zero (non-zero) quantum coherence initially.
\end{abstract}
\maketitle

\section{Introduction}\label{sec:Introduction}
 The pioneering works of Feynman \cite{feynman1986quantum, feynman1982simulating}, Bennett and Brassard \cite{BB84}, Deutsch \cite{deutsch1985quantum}, Shor \cite{shor1994algorithms}, and others have strongly established that using quantum resources in computation, communication, and metrology, one can obtain certain advantages usually referred to as \textit{quantum advantage}. Subsequent theoretical and experimental works including Google's recent experimental demonstration of quantum advantage \cite{arute2019quantum} have further established this fact (\cite{nagali2012experimental, kumar2019experimental, centrone2021experimental, bravyi2020quantum, novo2021quantum, maslov2021quantum} are a few recent examples). A common feature of all these works is that quantum advantage requires nonclassical states, or in other words, the quantum states having no classical analog. 
Nonclassical states may be generated by matter-field interactions. The Jaynes-Cummings model \cite{shore1993jaynes}, introduced in 1963 \cite{jaynes1963comparison}, describes one such simple interaction between a monochromatic field and a two-level atom.

This simple system, in the original form, as well as its various generalizations (see \cite{larsonspecial} in this context), is still relevant in several aspects of quantum information and QED. For instance, it plays a significant role in the generation of engineered quantum states \cite{dell2006multiphoton}, and the study of non-Markovian evolution \cite{bellomo2007non}. In fact, it is a convenient toy model---which has recently been generalized from different perspectives \cite{ermann2020jaynes, ghoshal2020population, villas2019multiphoton, huang2020ultrastrong}---often employed as a testbed to study various quantum effects, such as generation of Schr\"odinger-cat states \cite{buzek1992schrodinger}, entanglement protection \cite{fasihi2019entanglement}, catalysis \cite{Messinger_2020}, multiphoton blockade \cite{zou2020multiphoton}, fractional revivals \cite{averbukh1992fractional}, quantum state engineering \cite{gea1990collapse}, strong squeezing \cite{kuklinski1988squeezing}, entanglement generation \cite{phoenix1991entangled}, state discrimination \cite{namkung2019almost}.

In the usual Jaynes-Cummings model there are only two modes (atom mode and field mode), so relevant nonclassicality may appear either as two types of single-mode nonclassicalities involving each of these two modes, or as two-mode nonclassicality reflected as atom-field correlation.  Atom-field correlations in the Jaynes-Cummings model have already been studied using negativity \cite{akhtarshenas2007negativity}. Further, using the Wigner-Yanase skew information-based quantifier of nonclassicality described in \cite{luo2019quantifying}, single-mode nonclassicalities present in Jaynes-Cummings model have been studied \cite{fu2021dynamics, dai2020atomic}. Specifically, field nonclassicality was studied in \cite{fu2021dynamics}, and atomic nonclassicality in \cite{dai2020atomic}. Additionally, entanglement dynamics and the field nonclassicality in the double Jaynes-Cummings model are reported in \cite{ghorbani2017wigner}. Specifically, the correlations between the two atoms in the two-mode nonlinear Jaynes-Cummings model were quantified using von Neumann entropy, negativity, and concurrence, while two-mode field nonclassicality was witnessed by the Wigner quasiprobability distribution. The Wigner function neither quantifies the nonclassicality, nor captures all the nonclassicality of states; for example, it fails in the case of squeezed states \cite{agarwal2012quantum}. 

Thus, discrete efforts to study nonclassical features associated with the Jaynes-Cummings model have been made, but no attempt has yet been made either to look at the interplay between all three different aspects of nonclassicality which may be present in the Jaynes-Cummings model or to quantify the atom-field correlation, atomic nonclassicality, and field nonclassicality on the same footing. 
Here, we aim to do so and study atom-field correlations, atomic nonclassicality, and field nonclassicality using the same measure in the resonant Jaynes-Cummings model. Specifically, we use negativity potential (entanglement potential \cite{asboth2005computable}) to quantify the local nonclassicalities of the atom and field and negativity to measure atom-field correlation. Our main motivation is to understand the dynamics and interplay between single-mode and two-mode nonclassicalities observed in the Jaynes-Cummings model and also to verify whether the dynamics conserve total nonclassicality (\cite{ge2015conservation} and references therein). We consider distinct initial conditions which include four cases: (A) atom in the excited state and field vacuum, and atom in ground state with field in (B) Fock, (C) thermal, and (D) coherent state. The significant feature of Cases B-D is that Fock state are known to be most nonclassical state, while both mixture and superposition of Fock states cannot generate entanglement at a beam splitter. Further, coherent state has non-zero coherence in the Fock basis, unlike Fock and thermal state.

The rest of the paper is organized as follows. In Section \ref{sec:nonclassicality}, we discuss nonclassicality measures in general with a specific focus on the measures used in the present work.  In Section \ref{sec:dynamics}, we analyze the dynamics of nonclassicalities in detail for various initial conditions with specific attention to the question: Is total nonclassicality in the Jaynes-Cummings model conserved? Finally, the paper is concluded in Section  \ref{sec:conclusion}.

\section{Nonclassicality Measures}\label{sec:nonclassicality}

A nonclassical state cannot be represented as a statistical mixture of coherent states, and thus has a non-positive Glauber-Sudarshan $P$ function. This gives us the necessary and sufficient criterion of nonclassicality. However, this is not easy to compute for most of the states and does not quantify the amount of nonclassicality. Inspired by this, several measures of nonclassicality were proposed over the years, such as nonclassical distance (the distance from the nearest classical states) \cite{hillery1987nonclassical}, nonclassical depth (the amount of noise required to destroy the nonclassicality) \cite{lee1991measure}, nonclassical volume (volume of the negative part of the Wigner function) \cite{kenfack2004negativity}. However, these measures have some inherent limitations \cite{miranowicz2015statistical}. Nonclassical distance requires minimization over all possible classical states, nonclassical depth is unity for all pure non-Gaussian states \cite{lutkenhaus1995nonclassical}, and nonclassical volume fails for some squeezed states as Wigner function is non-negative. Further, entanglement potential \cite{asboth2005computable} was proposed by Asboth to quantify the {amount} of single-mode nonclassicality. It is the amount of entanglement at the output of an auxiliary beam splitter mixing the single-mode nonclassicality with any classical state. This allows us to use the same entanglement measure to quantify local atom and field nonclassicality (as entanglement potential) as well as atom-field correlations.

Specifically, we use negativity---introduced in \cite{zyczkowski1998volume} and shown to be LOCC monotone in \cite{vidal2002computable}---to quantify atom-field correlations (as well as entanglement potential). Negativity in a bipartite entangled state $\rho$ is the sum of the absolute values of the negative eigenvalues ($\lambda_k$) of the partial transposed density matrix $\rho^{\Gamma}$: 
\begin{equation}
    N(\rho) = \sum_{k: \lambda_k < 0} \left| \lambda_k \right| = \sum_{k} \frac{1}{2}\left(\left| \lambda_k\right| - \lambda_k\right).
\end{equation} 
This measure of entanglement is based on the partial transpose condition introduced by Peres \cite{peres1996separability}, which states that a bipartite density matrix is separable (classical) if its partial transpose has only non-negative eigenvalues. This is a necessary condition only for qubit-qubit and qubit-qutrit systems \cite{horodecki2001separability}. We study negativity here because it is easy to compute. As most entanglement measures involve an extremization, usually over complicated regions of the Hilbert space \cite{horodecki2009quantum}, computing the time evolution of such measures is difficult. In addition, negativity has a physical interpretation as the approximate number of entangled degrees of freedom in a bipartite system \cite{eltschka2013negativity}.

We use negativity potential as used in \cite{miranowicz2015statistical} to quantify nonclassicality in a single qubit. An analogous theoretical construction for the atomic states that defines a negativity potential for the atom \cite{naikoo2019interplay} is used here to quantify atomic nonclassicality. For brevity, we omit 'potential' in what follows and call them field negativity and atom negativity. 
The choice of negativity as a measure of entanglement is by no means unique, we could have chosen concurrence or logarithmic negativity instead (\cite{horodecki2009quantum} reviews these in detail). 
Further, Asboth's idea of entanglement potential is critically analyzed as it is shown that there exists some residual nonclassicality in the single-mode states at the output of the beam splitter which is not captured by entanglement \cite{ge2015conservation, arkhipov2016nonclassicality}. We also verify this in our context by using additional beam splitter layers to \emph{deplete} all the residual nonclassicality.

\section{Dynamics of nonclassiality in the resonant Jaynes-Cummings model\label{sec:dynamics}}

To study the interplay between different types of nonclassicality present in Jaynes-Cummings model, we first need to briefly describe the model.

In a conventional Jaynes-Cummings model, a two-level atom interacts with a single-mode quantized electromagnetic field. In what follows, we denote the field (atom) mode with a subscript $f$ ($a$) and the excited (ground) state of the atom  by $\ket{1_a}$  ($\ket{0_a}$). Under this notation, resonant Jaynes-Cummings Hamiltonian can be expressed as \cite{jaynes1963comparison}
\begin{equation}
   H = \hbar \omega N_f + \hbar\omega \sigma_3 + \hbar\lambda \left( \sigma_+ a_f + \sigma_- a_f^{\dagger}\right), 
\label{eq:H}\end{equation}
where $N_f = a_f^{\dagger}a_f$ is the number operator of the electromagnetic field-mode with frequency $\omega$, $\sigma_3 = \left(\ket{1_a}\bra{1_a} - \ket{0_a}\bra{0_a}\right)$, $\sigma_+ = \ket{1_a}\bra{0_a}$, $\sigma_- = \ket{0_a}\bra{1_a}$, and $\lambda$ is a coupling constant. The first two terms in (\ref{eq:H}) are the free Hamiltonians of a single-mode electromagnetic field and a two-level atom. Due to the interaction between atom and field, the atom may drop (jump) from the excited (ground) to ground (excited) state by emitting (absorbing) a photon. Neglecting the effect of ambient environment, we can obtain the unitary dynamics of an arbitrary initial atom-field state in the Heisenberg picture.
Interestingly, different initial states will lead to different dynamics of nonclassicality. Here, we discuss the dynamics of nonclassicality for a set of initial states. Specifically, we will provide the analytic expressions of both local nonclassicality and correlations for two initial states, which will be followed by some numerical results for two relatively complex initial states. To begin with we will discuss a simple case where the composite atom-field initial state is $\ket{0_f}\ket{1_a} $, implying that the filed is in vacuum and the atom is in excited state.

\subsection{Case A: Initial Field  in Vacuum and Atom in  Excited State }\label{sec:A Simple Case}

Under Jaynes-Cummings Hamiltonian (\ref{eq:H}), initial state $\ket{0_f}\ket{1_a}$ evolves as (see \cite{gerry2005introductory} for instance) 
\begin{equation}\label{01dynamics}
    \ket{0_f}\ket{1_a} \to \cos(\lambda t) \ket{0_f}\ket{1_a} - i \sin(\lambda t) \ket{1_f}\ket{0_a}.
\end{equation}
Thus, time evolution of the corresponding density matrix is
\begin{equation}\label{01dmatrix}
\begin{split}
    \rho(T)  =& \cos^2(T) \ket{0_f,1_a}\bra{0_f,1_a}+ \sin^2(T) \ket{1_f,0_a}\bra{1_f,0_a} \\ 
    & -\left(i \sin(T)\cos(T) \ket{1_f,0_a}\bra{0_f,1_a} + \text{H.c.}\right),
\end{split}
\end{equation}
where $\text{H.c.}$ stands for Hermitian conjugate, and $T = \lambda t$ is a dimensionless time parameter. 
Using this we can write the dynamics of the reduced density matrices for the field and atom as
\begin{eqnarray}
\rho_f(T) = \text{Tr}_{a} \left(\rho(T) \right)= \cos^2(T) \ket{0_f}\bra{0_f} + \sin^2(T) \ket{1_f} \bra{1_f}, \label{01fdmatrix}\\ 
\rho_a(T) = \text{Tr}_{f} \left(\rho(T) \right)=  \sin^2(T) \ket{0_a}\bra{0_a} + \cos^2(T) \ket{1_a}\bra{1_a}, \label{01admatrix}
\end{eqnarray}
by partial tracing the atom and field modes, respectively.

Let us now compute the analytic expressions of negativity to quantify the nonclassicalities present in the composite state $\rho(T)$ and the reduced single-mode states $\rho_{f}(T)$ and $\rho_{a}(a)$.

\subsubsection{Atom-Field Correlations} 
We can obtain the partial transposed density matrix from (\ref{01dmatrix}) as 
\begin{equation}\label{01dmatrixpt}
\begin{split}
    \rho^{\Gamma}(T)  =&  \cos^2(T) \ket{0_f,1_a}\bra{0_f,1_a}+ \sin^2(T) \ket{1_f,0_a}\bra{1_f,0_a}  \\ 
    & + \left(i \sin(T)\cos(T) \ket{0_f,0_a}\bra{1_f,1_a} + \text{H.c.}\right).
\end{split}
\end{equation}
As we know that partial transpose with respect to the atom or field gives the same eigenvalues and hence the same negativity.
The eigenvalues of the partial transposed density matrix are
\begin{equation}\label{eigc}
   \lambda^{(c)}_1 = \cos^2(T), \quad \lambda^{(c)}_2 = \sin^2(T),\quad \lambda^{(c)}_{3} = - \lambda^{(c)}_4 = \frac{1}{2}|\sin(2T)|.
\end{equation}
As only $\lambda_4^{(c)}$ is negative,  a closed form analytic expression for the atom-field negativity can be obtained as 
\begin{equation}\label{Nc}
    N_c(T) = |\lambda^{(c)}_4| = \frac{1}{2}|\sin(2T)|.
\end{equation}
If $N_c(T)=0$, the state is separable.

\subsubsection{Field Negativity}
The reduced density matrix of the field is given in Eq. (\ref{01fdmatrix}). If we inject it through one port of a symmetric beam splitter and an auxiliary vacuum $\ket{0_{f_0}}$ through the other port, the output would be 
\begin{equation}\label{rfbs}
    \rho_{f_{BS}}(T) = U_{BS} \left(\rho_f(T) \otimes \ket{0_{f_0}}\bra{0_{f_0}}\right) U_{BS}^{\dagger},
\end{equation}
where $U_{BS}$ is the unitary action of the beam splitter \cite{prasad1987quantum,thapliyal2019optical} and ${f_0}$ stands for the auxiliary bosonic field. 
We can compute  $\rho_{f_{BS}}(T)$  as
\begin{equation}\label{fbs}
\begin{split}
    \rho_{f_{BS}}(T)  = & \cos^2(T) \ket{0_f,0_{f_0}}\bra{0_f,0_{f_0}} + \frac{1}{2}\sin^2(T) \left(\ket{1_f,0_{f_0}}\bra{1_f,0_{f_0}} + \ket{0_f,1_{f_0}}\bra{0_f,1_{f_0}}\right) \\ 
    & -\frac{1}{2}\left({i}\sin^2(T)\ket{0_f,1_{f_0}}\bra{1_f,0_{f_0}} + \text{H.c.}\right) 
\end{split}
\end{equation}
{using the properties of beam splitter operation in Schr{\"o}dinger picture} 
\begin{equation}
\begin{split}
    U_{BS} \ket{1_f}\ket{0_{f_0}} =& \frac{1}{\sqrt{2}}\left(\ket{1_f,0_{f_0}} - i \ket{0_f,1_{f_0}}\right).
    \end{split}
\end{equation}
Also, the beam splitter operation does not change the vacuum input states $U_{BS}\ket{0_f}\ket{0_{f_0}} = \ket{0_f}\ket{0_{f_0}}$.

Partial transpose of density matrix (\ref{fbs}) with respect to modes $b$ yields 
\begin{equation}\label{fbspt}
\begin{split}
    \rho_{f_{BS}}^{\Gamma}(T) &= \cos^2(T) \ket{0_f,0_{f_0}}\bra{0_f,0_{f_0}} +  \frac{1}{2}\sin^2(T) \left(\ket{1_f,0_{f_0}}\bra{1_f,0_{f_0}} + \ket{0_f,1_{f_0}}\bra{0_f,1_{f_0}}\right)  \\ 
    &- \frac{1}{2}\left( {i}\sin^2(T)\ket{0_f,0_{f_0}}\bra{1_f,1_{f_0}} + \text{H.c.}\right).
\end{split}
\end{equation}
We further compute its eigenvalues as
\begin{equation}\label{eigf}
\begin{split}
\lambda^{(f)}_1  = \lambda^{(f)}_2 = \frac{\sin ^2(T)}{2},\quad
\lambda^{(f)}_j  = \frac{\cos^2(T)}{2} -(-1)^j\chi(T),  
\end{split}
\end{equation}
where
$\chi(T) = \frac{1}{4}\sqrt{3+\cos(4T)}$ and $j = 3,4$. Thus, the field negativity is obtained as
\begin{equation}\label{Nf}
\begin{split}
  N_f(T)  = \left|\lambda^{(f)}_4\right|=-\frac{1}{4} \left(1+\cos (2 T)-\sqrt{3+\cos (4 T)}\right),
\end{split}
\end{equation}
as only $\lambda^{(f)}_4$ is negative.

\subsubsection{Atom Negativity}

To obtain atom negativity, we use a beam splitter type operation 
\begin{equation}
\begin{split}
    U_{BS} \ket{1_a}\ket{0_{a_0}} =& \frac{1}{\sqrt{2}}\left(\ket{1_a,0_{a_0}} - i \ket{0_a,1_{a_0}}\right)
    \end{split}
\end{equation}
with an auxiliary mode ${a_0}$, in the same way as used for the field mode. The beam splitter operation in context of mapping the nonclassicality of a quantum state to the entanglement of the output \cite{vogel2014unified} is often discussed for single qubit \cite{miranowicz2015statistical} and spin states  \cite{markham2003classicality}, too.
The atomic state after the action of the beam splitter, obtained in the same way as Eq. (\ref{fbs}) with the help of an auxiliary mode ${a_0}$, is  
\begin{equation}\label{abs}
   \rho_{a_{BS}}(T)   = \rho_{f_{BS}}\left(T\pm\frac{\pi}{2}\right), 
\end{equation}
where the modes ${f_0}$ and $f$ are substituted by ${a_0}$ and $a$, respectively.
Similarly, the partial transpose of density matrix (\ref{abs}) with respect to the auxiliary mode $c$ is obtained and its eigenvalues are 
\begin{equation}\label{eiga}
         \lambda^{(a)}_1 = \lambda^{(a)}_2 = \frac{\cos ^2(T)}{2}, \quad \lambda^{(a)}_{i} = \frac{\sin^2(T)}{2} -(-1)^j\chi(T), \quad \text{$j=3,4$.}
\end{equation}
This allows us to quantify the atom negativity in terms of entanglement potential as
\begin{equation}\label{Na}
   N_a(T) = \left|\lambda_4^{(a)}\right|=N_f\left(T\pm\frac{\pi}{2}\right).
\end{equation}

\begin{figure}
\includegraphics[scale=0.75]{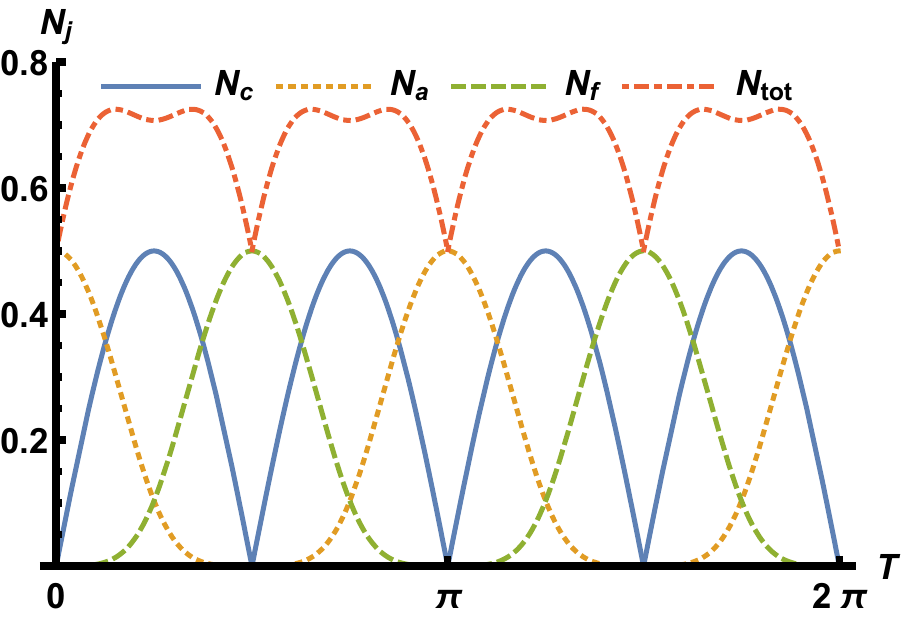} 

    \caption{(Color online) Atom-field correlations $N_{c}$, field negativity $N_{f}$, and atom negativity $N_{a}$, and total nonclassicality $N_{\rm tot}$ are shown as a function of the dimensionless time parameter $T=\lambda t$ for the field initially in vacuum and atom in the excited state.}
    \label{fig1}
\end{figure}

The analytic expressions describing the dynamics of the amount of atom-field correlation, and single-mode nonclassicalities present in the system (cf. Eqs. (\ref{Nc}), (\ref{Nf}), and (\ref{Na})) reveals that the periodic evolution of both the field and atom nonclassicality has a period $\pi$, while correlations have one-half the time period of that of local nonclassicalities. Dynamics of  atom-field correlation, field nonclassicality, and atom nonclassicality quantified via corresponding negativity are  plotted against $T$ in Fig. \ref{fig1}. It is observed that an exchange of nonclassicality between the field and atom is mediated through the correlations. Interestingly, when the correlations are zero, the atom and field negativities are alternatively maximum, which corresponds to the scenario when the nonclassicality is completely localized in the subsystems. Further, it is observed that even when the correlations are maximum the atom and field negativity potentials are non-zero. The origin of this interesting feature will be discussed later in Subsection \ref{sec:where}. 

To to see whether the total nonclassicality remains constant when the chosen initial state evolves under the Jaynes-Cummings Hamiltonian \ref{eq:H}, the total nonclassicality ($N_{\rm tot}(T)=N_c(T)+N_f(T)+N_a(T)$) is plotted in Fig. \ref{fig1}. Jaynes-Cummings dynamics is found to generate an additional nonclassicality as the total nonclassicality is found to be bounded from below by the initial amount of nonclassicality in the system, i.e., $N_{\rm tot}(T)\geq N_{a}(T=0)$.

\subsubsection{Residual Nonclassicalities}

To quantify the amount of nonclassicality in $\rho_j(T):\,j\in\{a,f\}$, we mapped it using a linear optical tool, a beam splitter as it cannot enhance the amount of nonclassicality ($\mathcal{N}$) \cite{vogel2014unified}, to the amount of entanglement in $\rho_{j_{BS}}(T)$. However, a more general systematic study should be the amount of nonclassicality $\mathcal{N}^{\rm in}$ in the input of the beam splitter should be equal to the amount of nonclassicality $\mathcal{N}^{\rm out}$ in the output of the beam splitter. We know that $\mathcal{N}^{\rm in}_j(T)=\mathcal{N} \left(\rho_j(T) \right)$ as the auxiliary mode ${j_0}$ is initially in vacuum ($\mathcal{N} \left(\rho_{j_0}(T) \right)=0$ and not correlated with $\rho_j(T)$. However,  $\mathcal{N}^{\rm out}_j(T)= \mathcal{E} \left(\rho_{j_{BS}}(T) \right)+\mathcal{N} \left(\rho_{j^{\prime}}(T) \right)+\mathcal{N} \left(\rho_{j_0^{\prime}}(T) \right)$, where $\rho_{j^{\prime}}(T)=\text{Tr}_{j_0} \left(\rho_{j_{BS}}(T) \right)$ and $\rho_{j_0^{\prime}}(T)=\text{Tr}_{j} \left(\rho_{j_{BS}}(T) \right)$ are the reduced single-mode states at the output of the beam splitter, and $\mathcal{E} \left(\rho_{j_{BS}}(T) \right) $ is the amount of entanglement in $\rho_{j_{BS}}(T)$ which is already calculated as $\mathcal{E} \left(\rho_{j_{BS}}(T) \right)= N_j(T)$. {In brief, we obtain the reduced states after the beam splitter as $\left\{\rho_j(T),\rho_{j_0}(T) \right\} \xrightarrow{U_{BS}} \left\{\rho_{j_{BS}}(T)  \right\}\xrightarrow{{\mathrm{Partial\, trace}}} \left\{\rho_{j^{\prime}}(T),\rho_{j_0^{\prime}}(T)  \right\}$.} 

We calculate the local nonclassicalities in the reduced single mode outputs of the beam splitter applied on the field mode $\mathcal{N} \left(\rho_{f^{\prime}}(T) \right)$ and 
$\mathcal{N} \left(\rho_{f_0^{\prime}}(T) \right)$ in terms of entanglement potential. Here, from Eq. (\ref{fbs}) we have
\begin{eqnarray}
   \rho_{f^{\prime}}(T)  = \left(\cos^2(T) + \frac{1}{2}\sin^2(T)\right) \ket{0_f}\bra{0_f} + \frac{1}{2}\sin^2(T)\ket{1_f}\bra{1_f}, \label{ff}\nonumber
   \\
  \rho_{f_0^{\prime}}(T) = \left(\cos^2(T) + \frac{1}{2}\sin^2(T)\right)\ket{0_{f_0}}\bra{0_{f_0}}  + \frac{1}{2}\sin^2(T)\ket{1_{f_0}}\bra{1_{f_0}}. \label{fb} 
\end{eqnarray}
This yields $ \rho_{f^{\prime}_{BS}}(T)$ the state after the beam splitter operation on $\rho_{f^{\prime}}(T)$, obtained in the same way as Eq. (\ref{fbs}) with the help of an auxiliary mode ${f_{10}}$, as \begin{equation}\label{ffbs}
\begin{split}
     \rho_{f^{\prime}_{BS}}(T) & = \left(\cos^2(T) + \frac{1}{2}\sin^2(T)\right)\ket{0_f,0_{f_{10}}}\bra{0_f,0_{f_{10}}}  
 + \frac{1}{4}\sin^2(T) \left(\ket{1_f,0_{f_{10}}}\bra{1_f,0_{f_{10}}} + \ket{0_f,1_{f_{10}}}\bra{0_f,1_{f_{10}}}\right) \\ 
    & -\frac{1}{4}\left({i}\sin^2(T)\ket{0_f,1_{f_{10}}}\bra{1_f,0_{f_{10}}} + \text{H.c.}\right). 
\end{split}
\end{equation}
It is worth noting here that $\rho_{f_0^{\prime}}(T)$ and $\rho_{f^{\prime}}(T)$ have the same form in Eq. (\ref{fb}). Consequently, the amount of nonclassicality will be the same. 
We compute the eigenvalues of the partial transpose of density matrices $\rho_{f_0^{\prime}}(T)$ and $\rho_{f^{\prime}}(T)$ as 
    \begin{equation}\label{eigff}
    \begin{split}
 \lambda^{(f_1)}_1  = \lambda^{(f_1)}_2 = \frac{\sin ^2(T)}{4},\quad
\lambda^{(f_1)}_j  =\frac{1}{8} \left(3+\cos (2 T) -(-1)^j\xi(T)\right),
    \end{split}
\end{equation}
where $\xi(T) = \sqrt{11+4 \cos (2 T)+\cos (4 T)}$ and $j = 3,4$. This gives us the nonclassicality in the states $\rho_{f_0^{\prime}}(T)$ and $\rho_{f^{\prime}}(T)$ as
\begin{equation}\label{Nff}
\begin{split}
  N_{f_1}(T)  = \left|\lambda^{({f_1})}_4\right|=-\frac{1}{8} \left(3+\cos (2 T)-\xi(T)\right),
\end{split}
\end{equation}
as only $\lambda^{(f_1)}_4$ is negative.

\begin{figure}
    \begin{centering}
\subfloat[]{\begin{centering}
\includegraphics[scale=0.4]{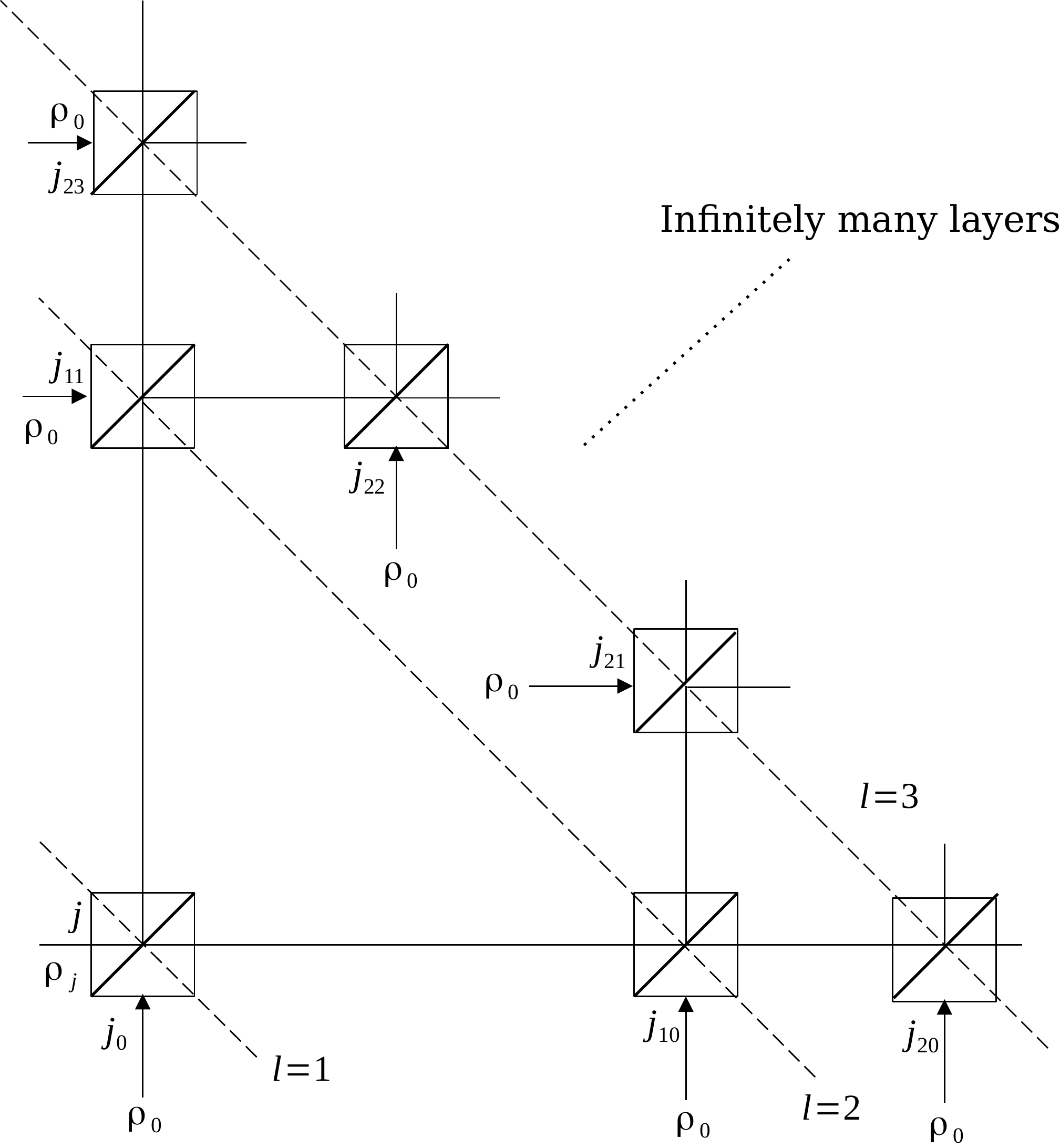}
\par\end{centering}

}\subfloat[]{\begin{centering}
\includegraphics[scale=0.75]{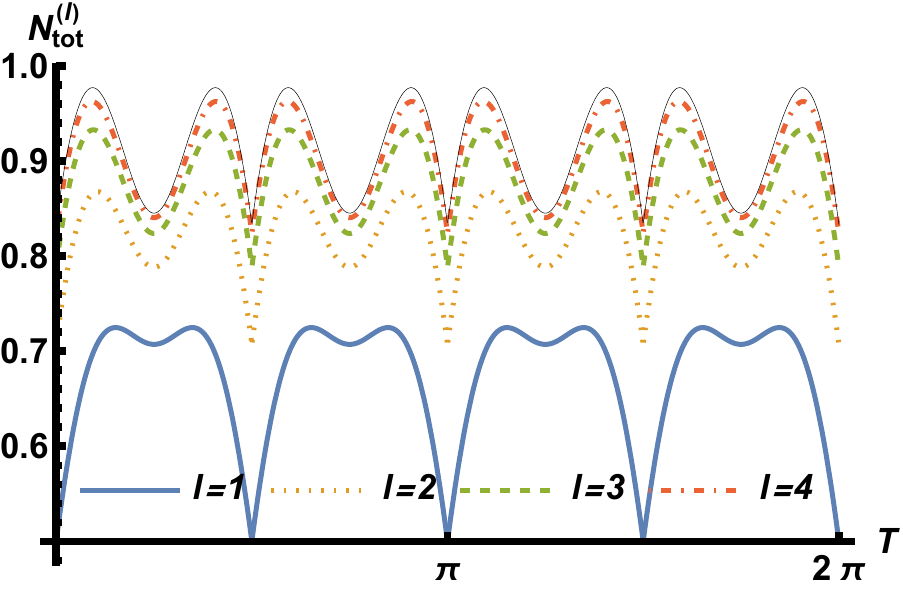}
\par\end{centering}
}
\par\end{centering}    
    \caption{(Color online) (a) Additional layers of beam splitter used to quantify the nonclassicality and to deplete the residual nonclassicality. Here, $\rho_0$ denotes the density matrix of single mode vacuum and $j\in\{a,f\}$. (b) Total amount of nonclassicality $N_{\rm tot}^{(l)}$ quantified after $l$ layers of beam splitters. The thin (black) line corresponds to $N_{\rm tot}^{(\infty)}$ in (b). }
    \label{fig3}
\end{figure}

Interestingly, $N_{f_1}(T)\neq0$ for an arbitrary value of $T$ leads us to conclude that $\mathcal{N}^{\rm out}_f(T)\geq N_f(T)$ as we have obtained $\mathcal{N} \left(\rho_{j^{\prime}}(T) \right)=\mathcal{N} \left(\rho_{j_0^{\prime}}(T) \right)=N_{f_1}(T)$. We call this nonclassicality in the reduced single mode output states as residual nonclassicality \cite{ge2015conservation}. Adopting the similar mechanism for the reduced single mode outputs of the beam splitter applied on the atomic state we obtain $\mathcal{N} \left(\rho_{a^{\prime}}(T) \right)$, which is the same as $\mathcal{N} \left(\rho_{a_0^{\prime}}(T) \right)$, in terms of entanglement potential which gives us 
\begin{equation}\label{Naa}
  N_{a_1}(T)  =N_{f_1}\left(T\pm\frac{\pi}{2}\right).
\end{equation}
However, the result so far allow us to conclude that $\mathcal{N}^{\rm out}_j(T)\geq N_j(T)+2 N_{j_1}(T)$ as we cannot discard the residual nonclassicality in the reduced single mode states of $\rho_{j^{\prime}_{BS}}(T)$ and the rest of the modes. For this reason, we can add a new layer of beam splitters as shown in Fig. \ref{fig3} (a) until the local nonclassicality in the outputs of the beam splitters seizes.

Further, the total nonclassicality after two layers of beam splitters can be written as 
\begin{equation}\label{N2}
    N_{\rm tot}^{(2)}(T) = N_c (T)+ N_f (T)+ N_a(T) + 2 N_{f_1}(T) + 2 N_{a_1}(T).
\end{equation}
The variation of $N_{\rm tot}^{(2)}(T)$ in Fig. \ref{fig3} (b) clearly shows that $N_{\rm tot}^{(2)}(T)>N_{\rm tot}(T)\,\forall T$. This motivated us to obtain the total nonclassicality $N_{\rm tot}^{(l)}(T)$ obtained as sum of nonclassicality after $l$ layers of beam splitters. We obtained the analytic expressions for $N_{\rm tot}^{(l)}(T)$ upto $l=4$ and reported the variation in Fig. \ref{fig3} (b). We can observe that $N_{\rm tot}^{(l)}(T)> N_{\rm tot}^{(l-1)}(T)$ and the change, i.e., $N_{\rm tot}^{(l)}(T)-N_{\rm tot}^{(l-1)}(T)$, at the higher values of $l$ is less prominent as compared to that for the smaller number of s layers of beam splitter used.

We have already mentioned that due to symmetry of the system we observed that the amount of residual nonclassicality $N_{a_{l}}(T)\, \left(N_{f_{l}}(T) \right)$ in all the atomic (field) modes after $l+1$ layers of beam splitters is the same. Further, the amount of nonclassicality is also observed to follow $N_{a_{l}}(T)=N_{f_{l}}\left(T\pm\frac{\pi}{2} \right)$. We also observed that nonclassicality depletes approximately by a factor of one-fifth, i.e.,  $N_{j_{l}}(T)\approx5N_{j_{l-1}}(T)$.
Using this we can obtain the total nonclassicality (after an infinitely large number of beam splitter layers) approximately using the sum of geometric series as
\begin{equation}\label{Ninf}
    N_{\rm tot}^{(\infty)}(T) \approx N_c (T)+ \sum\limits_{n=0}^{\infty} \left(\frac{2}{5}\right)^n \left(N_f (T)+ N_a(T)\right) = N_c (T)+ \frac{5}{3}\left(N_f (T)+ N_a(T)\right). 
\end{equation}

\subsubsection{Origin of extra nonclassicality}\label{sec:where}

We have observed in Figs. \ref{fig1} and \ref{fig3} that $N_{\rm tot}^{(l)} (T)\geq N_{a}(T=0)\forall\, T,\, l$. Specifically, when the correlations are maximum $N_{c} \left(T=\frac{ (2 n+1)\pi}{4}\right)= N_{a}(T=0)\,:n\in \mathbb{Z}$, the atom and field negativity potentials were observed to be non-zero, i.e., $N_{j} \left(T=\frac{ (2 n+1)\pi}{4}\right)\neq 0:\,j\in\{a,f\}$. 
Interestingly, upon closer examination of the reduced states in Eqs. (\ref{01fdmatrix}) and (\ref{01admatrix}), we can observe that $\rho_f\left(T=\frac{ (2 n+1)\pi}{4}\right)=\rho_a\left(T=\frac{ (2 n+1)\pi}{4}\right)$ are maximally mixed states. The nonclassicality present in the maximally mixed states, obtained here as a statistical mixture of a classical (vacuum) and a nonclassical (single photon Fock) states with equal probability, can be attributed to the presence of highly nonclassical Fock state.  
The observed behavior is consistent with the earlier findings of some of the present authors \cite{miranowicz2015statistical} that statistical mixtures of vacuum and single photon states can be highly nonclassical. Thus, as long as $T\neq \frac{n\pi}{2}$, this contribution increases the total nonclassicality in the system.

In this subsection, we have extensively discussed the dynamics of nonclassicalities in the resonant Jaynes-Cummings model considering that the initial state is $ \ket{0_f}\ket{1_a} $. In the following subsections, we will concisely report how the dynamics of nonclassicality changes if we change the initial state.

\subsection{Case B: Initial Field in Fock State and Atom in Ground State}
In Case A, we have studied the case where initially the field was classical (vacuum) and the atom was excited. Here, we assume the field in the (nonclassical) Fock state $\ket{2_f}$ and the atom in the ground state initially. Under the Jaynes-Cummings dynamics, this state evolves as (see \cite{gerry2005introductory} for reference)

\begin{equation}\label{20dynamics}
    \ket{2_f}\ket{0_a} \to \cos(T\sqrt{3})\ket{2_f}\ket{0_a} - i \sin(T\sqrt{3}) \ket{1_f}\ket{1_a}. 
\end{equation}

\begin{figure}
    \begin{centering}
\subfloat[]{\begin{centering}
\includegraphics[scale=0.6]{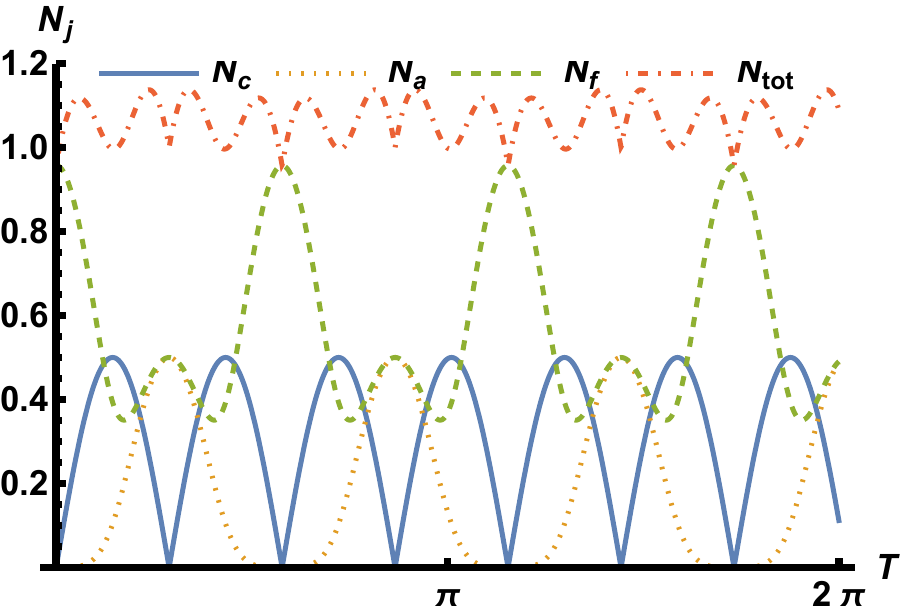}
\par\end{centering}

}\subfloat[]{\begin{centering}
\includegraphics[scale=0.6]{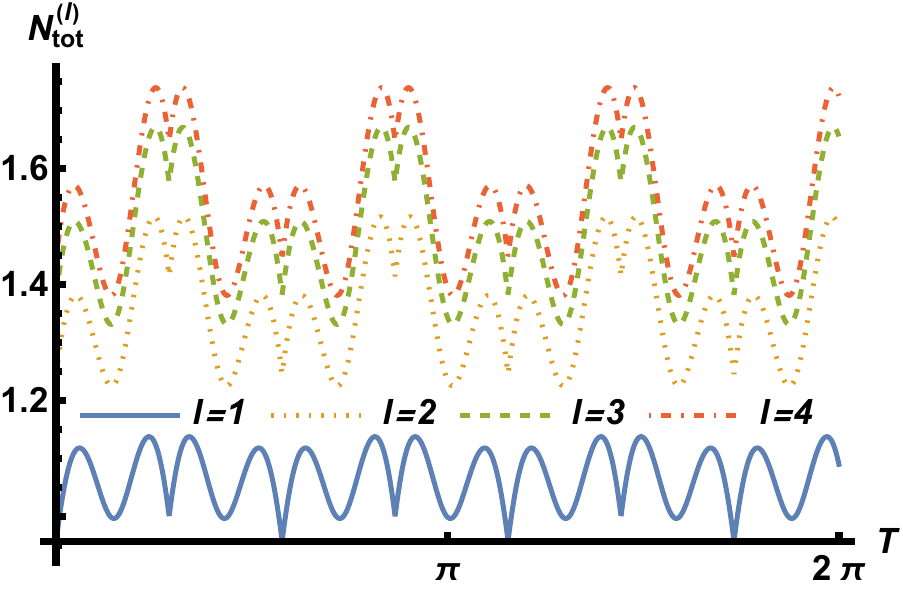}
\par\end{centering}
}
\par\end{centering}    
    \caption{(Color online) (a) Atom-field correlations, field negativity potential, and atom negativity potential as functions of $T$ for the initial state $\ket{2_f}\ket{0_a}$. (b) Total nonclassicality for the initial state $\ket{2_f}\ket{0_a}$ up to four layers of beam splitters.}
    \label{fig7}
\end{figure}

Computation of negativity is performed using similar strategy as was adopted for Case A. Specifically, we obtained the atom-field correlation and atom negativity as 
\begin{equation}\label{Nf-CB}
\begin{split}
   N_c(T)  &=\frac{1}{2} \left| \sin \left(2 \sqrt{3} T\right)\right|,\\
  N_a(T)  &= -\frac{1}{4} \left(1+\cos \left(2 \sqrt{3} T\right)-\sqrt{3+\cos \left(4 \sqrt{3} T\right)}\right),
\end{split}
\end{equation}
respectively. The analytic expression of the field negativity is not included here as the expression is  cumbersome and it's difficult to deduce any physical insight from that. We can clearly observe from Eq. (\ref{Nf-CB}) that both the negativity expressions have periodic behavior where $N_c(T)$ and $N_a(T)$ have period $\frac{\pi}{2\sqrt{3}}$ and $\frac{\pi}{\sqrt{3}}$, respectively.
The obtained results are plotted  in Fig. \ref{fig7} (a) which illustrates the evolution of the atom-field correlation, the field negativity, and the atom negativity in this case. Initially, all the nonclassicality resides in the field, in contrast to Case A. Further, we can observe that the total initial nonclassicality (which is nonclassicality in the field) in Case B is higher than that observed in Case A (which is nonclassicality in the atom). This is due to the fact that nonclassicality of Fock state $\ket{n}$ is known to increase with $n$. Interestingly, negativity of field has the same period as that of atom negativity, which is consistent with Case A. 
When the correlations are zero (for $T=\frac{m\pi}{2\sqrt{3}}$), two possibilities are observed: (i) the atom negativity is zero and the field negativity is maximum (for even values of $m$); corresponding to $\ket{2_f}\ket{0_a}$, and (ii) atom negativity is maximum and the field negativity is $0.5$ (for odd values of $m$); corresponding to $\ket{1_f}\ket{1_a}$. We again observe that nonclassicality is exchanged between the field and atom through the correlations. Also notice that $N_f\left(T=\frac{\left(2k+1\right)\pi}{2\sqrt{3}}\right)=0.5\neq0\,\forall\,k\in\mathbb{Z}$, which corresponds to the case when field negativity is maximum in Case A.

When the atom-field correlations are maximum (for $T=\frac{\left(2k+1\right)\pi}{4\sqrt{3}}\,\forall\,k\in\mathbb{Z}$), the field negativity and the atom negativity are non-zero. Specifically, in this case, the atom is in maximally mixed states of ground and excited state; whereas the field is in the statistical mixture of two nonclassical (single and two photon Fock) states with equal probability. Therefore, the nonclassicalicality in the field in this particular case is less than that observed for pure Fock states $\ket{n=1}$ and $\ket{n=2}$, which are $1/2$ and $(1+2\sqrt{2})/4$, respectively. 

We also computed the total nonclassicality $N_{\rm tot}^{(l)}(T)$ including the residual nonclassicality depleted using $l\leq 4$ layers of beam splitters. Variation of $N_{\rm tot}^{(l)}(T)$ is shown in Fig. \ref{fig7} (b). The observation here is consistent with that of Case A, i.e., total nonclassicality in the Jaynes-Cummings dynamics is bounded below by the amount of initial nonclassicality in the atom-field system. Further, $N_{\rm tot}^{(\infty)}$ obtained in Case A is not valid in this case due to larger dimension of the truncated Fock space corresponding to the field mode. 

We have considered so far both atom and field in the pure states. In what follows, we consider a case with field initially in the mixture of Fock states and other that with the superposition of Fock states. For the sake of comparative study, we assume both the states with the same average photon number.

\subsection{Case C: Initial Field in Thermal State and Atom in Excited State}

Once again we assume all the initial nonclassicality in the atom and consider the field to be in the thermal state initially, which is defined as 
\begin{equation}\label{rth}
\rho_{\rm th} = \sum_{n=0}^{\infty} p_ n \ket{n_f}\bra{n_f}=
    \sum_{n=0}^{\infty} \frac{\langle n \rangle^n}{\left(\langle n \rangle + 1\right)^{n+1}}\ket{n_f}\bra{n_f},
\end{equation}
where $\langle n \rangle$ is the average photon number of the thermal state.
We can consider a small value of average photon number to truncate the thermal state to a form as a statistical mixture of first few Fock states with the weight defined by $p_ n$ in Eq. (\ref{rth}). For instance, for $\langle n \rangle = 0.01$, $p_ 0=\frac{100}{101}$ and $p_ 1=\frac{100}{101^2}\approx1-p_0$ with $p_ j=0\,\forall\, j>1$, which allows us to truncate the thermal state at $n=2$. 
Thus, combined initial state of the atom and field is $\rho_{\rm th}\otimes \ket{1_a}\bra{1_a}$. 
We truncate the series at the second term to ensure that the field remains effectively in the three dimensional space ($\ket{0_f}, \ket{1_f}, \ket{2_f}$) under evolution which allows us to perform our study in qubit-qutrit system, where the partial transpose condition is both necessary and sufficient to detect entanglement \cite{horodecki2001separability}. Specifically, when the atom returns to ground state, it transfers the excitation to the field, which is now obtained in photon added thermal state \cite{agarwal1992nonclassical,zavatta2007experimental}. This is one of the significant methods of adding photons in arbitrary quantum states of field to obtain a desired state (see \cite{malpani2021can} and references therein for a detailed discussion).

\begin{figure}
    \begin{centering}
\subfloat[]{\begin{centering}
\includegraphics[scale=0.6]{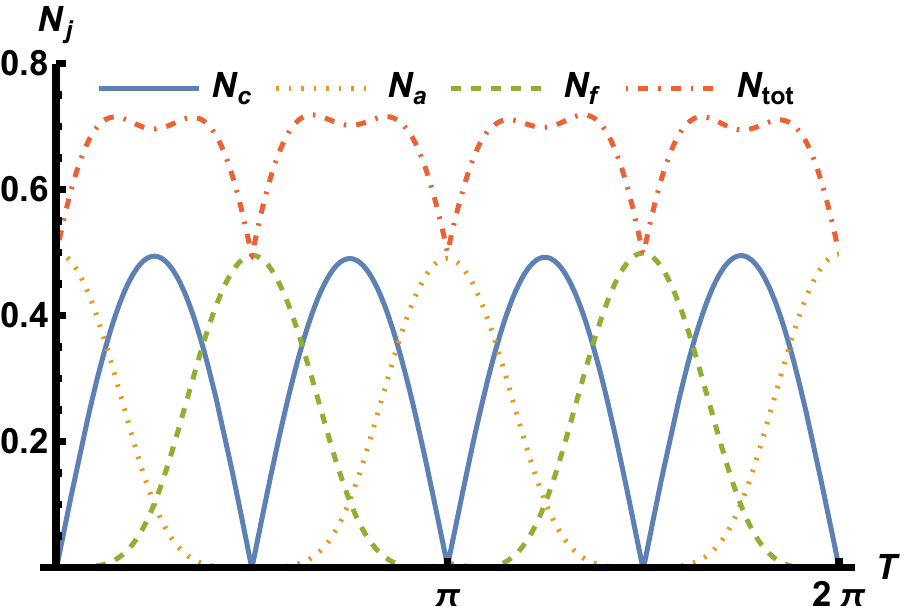}
\par\end{centering}

}\subfloat[]{\begin{centering}
\includegraphics[scale=0.6]{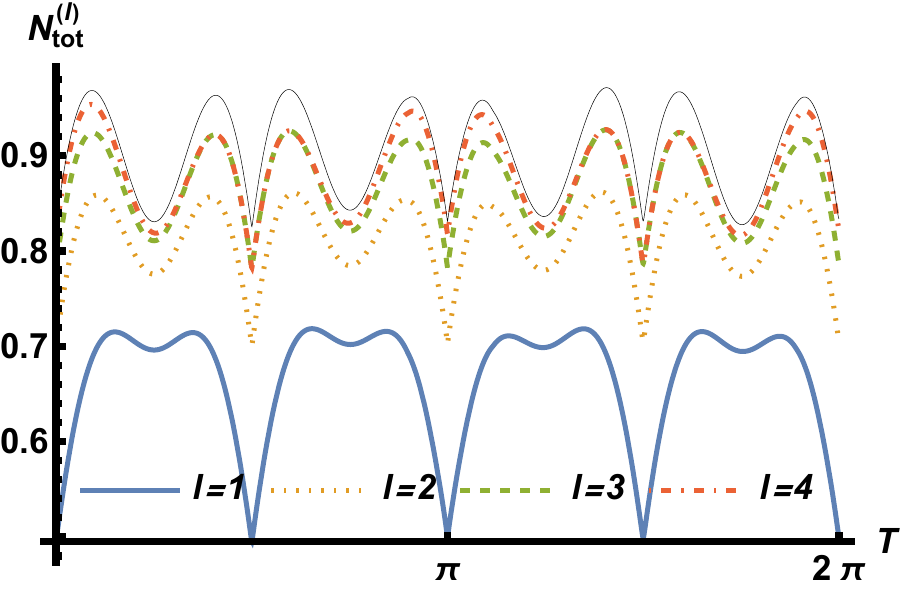}
\par\end{centering}
}
\par\end{centering}    
    \caption{(Color online) (a) Atom-field correlations, field negativity, and atom negativity as functions of $T$ for field initially in thermal state with average photon number $\langle n \rangle = 0.1$ and atom in the excited state. (b) Total nonclassicality $N_{\rm tot}^{(l)}(T)$ up to $l=4$ layers of beam splitters in this case.}
    \label{fig9}
\end{figure}

We can observe that the variation in Fig. \ref{fig9} (a) is similar to Fig. \ref{fig1},
as $p_1\ll p_0$. It can be easily verified from the reduced atom and field density matrices which are computed as
\begin{eqnarray}
   \rho_{a}(T)  = \left(p_0 \sin^2(T) +p_1 \sin^2(\sqrt{2}T)\right) \ket{0_a}\bra{0_a} + \left(p_0 \cos^2(T) +p_1 \cos^2(\sqrt{2}T)\right) \ket{1_a}\bra{1_a},\nonumber
   \\
  \rho_{f}(T) =p_0 \cos^2(T)\ket{0_f}\bra{0_f}+ \left(p_0 \sin^2(T) +p_1 \cos^2(\sqrt{2}T)\right) \ket{1_f}\bra{1_f} + p_1 \sin^2(\sqrt{2}T)\ket{2_f}\bra{2_f}. \label{fb-ther} 
\end{eqnarray}
For relatively higher values of average photon number (corresponding to higher values of $p_1$ but $p_2\approx 0$), we can observe that unlike in Case A the curves were not periodic due to contribution from both the cases but the qualitative behaviour remained identical. Specifically, the atom and field exchange nonclassicality through the correlations. When the correlations are maximum, the other two remain equal and non-zero. 

We also computed the total nonclassicality $N_{\rm tot}^{(l)}(T)$ up to $l=4$ layers of  beam splitters in this case, too. We have shown the variation of total nonclassicality in Fig. \ref{fig9} (b), where $N_{\rm tot}^{(\infty)}$ is also illustrated. 
 
\sloppy Interestingly, when the entanglement is maximum, the atomic state (from Eq. (\ref{fb-ther})) can be written as ~$p_0 \left(\ket{0_a}\bra{0_a}+\ket{1_a}\bra{1_a}\right) +p_1 \left(\sin^2(\frac{\pi}{2\sqrt{2}})\ket{0_a}\bra{0_a} +\cos^2(\frac{\pi}{2\sqrt{2}})\ket{1_a}\bra{1_a}\right).$ Therefore, the observed behavior, i.e., the total nonclassicality is not conserved in this case, can be inferred in the same way as in Case A in the limits of $p_1\ll p_0$. Further, we can observe in Eq. (\ref{fb-ther}) that when field nonclassicality is maximized a hole is burned in the photon number distribution at vacuum, which is a characteristic feature of photon added thermal state.

 \subsection{Case D: Initial Field in Coherent State and Atom in Excited State}
 
Yet another classical state with the distinct feature that it is the only pure classical state is a coherent state, which is defined as
 \begin{equation}\label{coherent}
 \ket{\alpha} =\sum_{n=0}^{\infty} c_n \ket{n}= e^{-\frac{1}{2}|\alpha|^2}\sum_{n=0}^{\infty} \frac{\alpha^n}{\sqrt{n!}} \ket{n}.
 \end{equation}
 Similar to the thermal state case, we  choose a small value for $\alpha$ such that $\langle n \rangle =\left|\alpha\right|^2= 0.01$ so that we may truncate the infinite series at $n=2$. Specifically, we chose $\alpha = 0.1$ and so the initial field has $c_0=e^{-\frac{1}{200}}$ and $c_1=\frac{c_0}{10}$ with $c_2\approx 0$
in Eq. (\ref{coherent}). 
When coupled with the excited atom, the field spans an effective three-dimensional space and the partial transpose condition becomes necessary and sufficient. 

\begin{figure}
\includegraphics[scale=0.6]{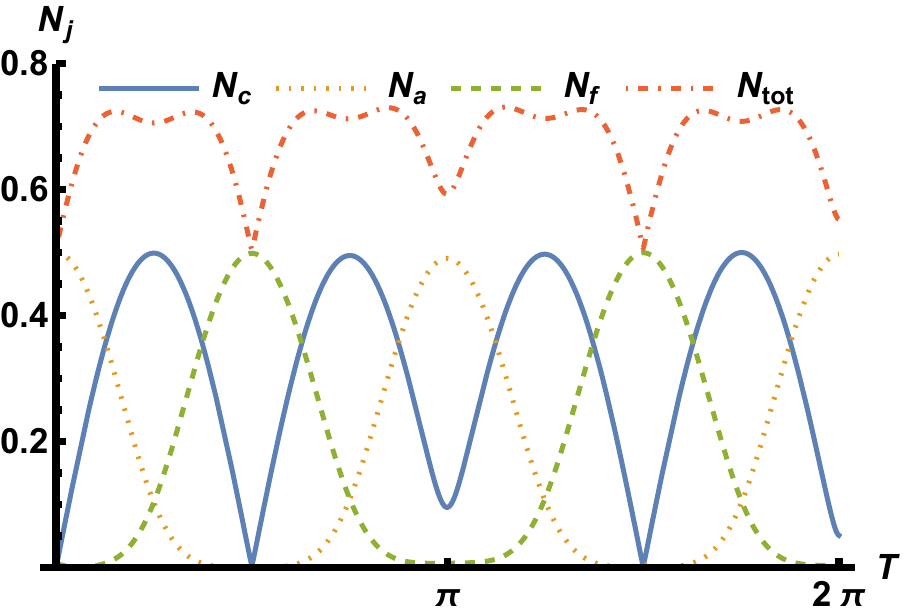}
    \caption{(Color online) Atom-field correlations, field negativity, atom negativity, and total nonclassicality for the field initially in coherent state $\ket{\alpha = 0.1}$ and atom in the excited state.}
    \label{fig11}
\end{figure}

Dynamics of quantified nonclassicality in this case is obtained numerically and the same is illustrated in Fig. \ref{fig11}.  In Fig. \ref{fig11}, we again observe the same dynamics of nonclassiality  as we have been observing for the other initial states. When the atom-field correlations are minimum, the field and the atom nonclassicalities are alternatively maximum. When the atom-field correlations are maximum, the field and atom negativity potentials are equal and non-zero. Nonclassicality is exchanged between the field and atom via correlations. However, Cases A-C correspond to cases where the reduced atomic and field states are always mixed state, i.e., all the quantum coherence present in the state is in the bipartite entangled state only. In contrast, in the present case, the  initial field has non-zero coherence which is transferred to atom as well.   This fact can be easily observed through the analytic expressions of the reduced atom and field density matrices which are computed as
\begin{equation}
\begin{split}
\rho_a(T) = & \left(c_0^2 \sin^2(T) + c_1^2 \sin^2(\sqrt{2}T)\right)\ket{0_a}\bra{0_a} + \left( i c_0 c_1 \cos(\sqrt{2}T)\sin(T)\ket{1_a}\bra{0_a} + \text{H.c.}\right) \\ & + \left(c_0^2\cos^2(T) + c_1^2 \cos^2(\sqrt{2}T)\right)\ket{1_a}\bra{1_a}, \\
\rho_f(T) & = c_0^2\cos^2(T)\ket{0_f}\bra{0_f} + \left(c_0c_1\cos(T)\cos(\sqrt{2}T)\ket{0_f}\bra{1_f} + \text{H.c}\right) + \left(c_0^2\sin^2(T) + c_1^2\cos^2(\sqrt{2}T)\right)\ket{1_f}\bra{1_f} \\ & + \left(c_0c_1 \sin(T)\sin(\sqrt{2}T)\ket{1_f}\bra{2_f} + \text{H.c}\right) + c_1^2\sin^2(\sqrt{2}T)\ket{2_f}\bra{2_f}.
\end{split} \label{eq:reduced-D}
\end{equation}
It can be observed from the above expressions of the reduced atom and field states that in the absence of the off-diagonal terms Eq. (\ref{eq:reduced-D}) reduces to  Eq. (\ref{fb-ther}) as $p_j\equiv c_j^2$. The distinct features observed in Fig. \ref{fig11} may be attributed to the quantum coherence in the subsystems. For instance, the reduced atomic state has non-zero quantum coherence in the Fock basis when the atom-field correlation is maximum, i.e., at $T=\pi/4$. Further, analogous to the photon added thermal state, the maximum nonclassicality in the field is obtained when the field is in photon added coherent state. The maximum and minimum of atom and field negativities as well as correlations can be observed at the similar conditions as in Case A-C because at lower average photon number vacuum term plays more significant role than any other non-zero photon term.

The total nonclassicality after one layer of beam splitters is illustrated in Fig. \ref{fig11}. Due to the computational difficulty of the problem we were unable to proceed beyond one beam splitter layer. We observed the Jaynes-Cummings dynamics failing to conserve the amount of nonclassicality in this case due to truncation of coherent state.

For all the different initial states studied here, we have observed residual nonclassicalities and the fact that nonclassicality is not conserved (total nonclassicality is not constant) under Jaynes-Cummings dynamics.

\section{conclusion\label{sec:conclusion}}

Nonclassical features associated with a resonant Jaynes-Cummings model (e.g.,  atom-field correlations, atomic nonclassicality, and field nonclassicality) have been studied for a set of  initial states to reveal the interplay between these three different aspects of nonclassicality. The analysis is performed using negativity potential (for single-mode nonclassicality) and negativity (for atom-field correlation), where atom and field negativity are computed using Asboth's criterion which tells us that if a single-mode nonclassical state is mixed with vacuum at a beam splitter then the output state becomes entangled, and the amount of entanglement present in the output state can be used as a quantitative measure of the input single-mode nonclassicality.  An analysis of Asboth's criterion performed here has revealed that there exists some residual nonclassicality in the single-mode states at the output of the beam splitter which is not captured by entanglement. We have verified this  using additional layers of beam splitters  and tried to deplete all the residual nonclassicality. The investigation has revealed that almost all the residual nonclassicality is captured with three layers of beam splitters.  Further, the origin of residual nonclassicality has also been investigated and the same is attributed to the Jaynes-Cummings dynamics.

The analysis of  dynamics of nonclassicality in the Jaynes-Cummings model has also revealed that nonclassicality is exchanged between the field and atom through atom-field correlations. {Further, it is observed that  the reduced states of the atom and field are found to be a statistical mixture of vacuum and Fock states when the atom-field correlations are maximum and the initial states of atom and field have no quantum coherence. However non-zero quantum coherence (in the Fock basis) in the initial atom-field state is transferred to both reduced atom and field states when the atom-field correlation are maximum. In both these situations, when the two-mode nonclassicality (correlation) has a maximum value, the single-mode nonclassicalities do not vanish. Further, it is found that when the correlations vanish, the atom and field negativities are alternatively maximum, implying that the nonclassicality is completely localized in the subsystems in those situations.} These observations motivated us to address the question: Is total nonclassicality conserved under Jaynes-Cummings dynamics? The study has revealed that it is not strictly conserved, but it can be viewed as quasi-conserved. Specifically, we have observed that Jaynes-Cummings dynamics generates an additional nonclassicality and the total nonclassicality is observed to be bounded below by the initial amount of nonclassicality present in the system. In short, the study has revealed many new features associated with the dynamics of nonclassicality present in the Jaynes-Cummings model, and it is hoped that others interested in nonclassicality dynamics will find the approach developed and used here useful for their studies involving other models of matter-field interaction.

\section*{Acknowledgement}
SA thanks José Luis Gómez-Muñoz and Francisco Delgado for their  "Quantum" package for \textit{Mathematica}. It helped in the computations.  AP acknowledges the support from the QUEST scheme of Interdisciplinary Cyber Physical Systems (ICPS) program of the Department of Science and Technology (DST), India (Grant No.:
DST/ICPS/QuST/Theme-1/2019/14 (Q80)). KT acknowledges support from
ERDF/ESF project `Nanotechnologies for Future' (CZ.02.1.01/0.0/0.0/16 019/0000754).

\end{document}